\documentclass{vospwks2007}

\title{Spectroscopic Surveys: Present}
\author{Ching-Wa Yip}
\affil{Department of Physics and Astronomy, The Johns Hopkins
  University, Baltimore, MD 21218, USA}

\newcommand{\solarmass}{M$_{\odot}$}

\newcommand{\lsim}{\stackrel{<}{{}_{\sim}}}

\usepackage{graphicx}

\begin{document}


\maketitle

\begin{abstract}
 I summarize  the current  spectroscopic sky surveys  and some  of the
 scientific results,  emphasizing the largest sky  survey to-date, the
 Sloan  Digital  Sky  Survey  (SDSS).   Techniques  used  commonly  in
 spectral analyses  are discussed, followed  by the present  needs and
 challenges for solving  some of the unknown problems.   I discuss how
 the  Virtual Observatory (VO)  can help  astronomers in  carrying out
 related research.
\end{abstract}

\section{Introduction}

The  homogeneity and  completeness of  large datasets  in  sky surveys
favor  them  to be  powerful  platforms  for  scientific research  and
discovery.  Depending  on the types  of objects, the  observed spectra
are imprinted with the most  vital information on their distances from
us,  rotational   velocities,  chemical  compositions   and  formation
histories.   Spectroscopic   sky  surveys  have   been  invaluable  to
astronomers in all areas nowadays.

\begin{figure}
\begin{center}
\scalebox
{0.43}
{
\includegraphics{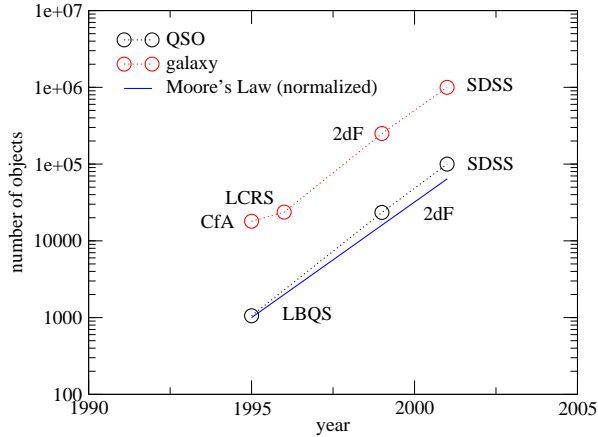}
}
\end{center}
\caption{The  number  of  spectroscopic  redshifts measured  from  sky
surveys  after   the  1990s,  for  galaxies  (red)   and  Type-I  AGNs
(black).  The  rate  of   increase  follows  nearly  the  Moore's  Law
(blue). (LBQS:  the Large  Bright Quasar Survey;  2dF: the  two degree
field survey; CfA: the  Center for Astrophysics redshift survey; LCRS:
the Las Campanas redshift survey)}
\label{fig:znumber}
\end{figure}

Starting  from  the 1990s,  the  rate of  increase  of  the number  of
spectroscopic     redshifts      measured     from     sky     surveys
(Fig.~\ref{fig:znumber})    follows    nearly    the    Moore's    Law
\citep{MooresLaw}. While  this coincidence is probably  by the designs
of  the surveys, the  speed of  increase of  the data  is high  by the
common  standard in  computer  science.   Not to  be  confined to  the
extra-galactic sky  surveys, large samples of stellar  spectra are now
being   taken  in  several   sky  surveys   aiming  at   producing  an
unprecedented view of our Milky  Way galaxy.  The sheer amount of data
(e.g.,  $\approx{\rm  TByte}$  of  data  in  the  SDSS)  promises  new
scientific   discoveries  across  diverse   areas  in   astronomy,  it
nonetheless presents  great challenges to astronomers  in handling the
higher-dimensional  data and extracting  physical parameters. 

The  Virtual Observatory \citep[VO,][]{2001Sci...293.2037S}  answers a
lot of these challenges by allowing astronomers to access data via the
internet in  an unified  way, and providing  to them web  services for
performing  analyses  on  the  data.  Continuous  discussions  between
astronomers   and  VO   builders  are   important  for   ensuring  the
practicality of the tools.

This overview  is concerned with  some of the interesting  results and
discoveries from the recent  spectroscopic sky surveys; and the common
techniques used in spectral analyses. They in turn would be argued for
the  present needs  and challenges,  in  the scopes  of both  spectral
analyses and VO.

\section{Current Spectroscopic Surveys}

The    recent     spectroscopic    surveys    are     summarized    in
Table~\ref{tab:surveys}.    They  are   ground-based,   with  observed
wavelengths mostly in the optical,  and the number of spectra taken to
be $> 10^{4}$ (approximately the number of spectra taken by the LBQS).
Two characteristics  are immediately evident: the  most recent surveys
tend to  go deeper instead of  getting more redshifts; and  the use of
magnitude selection instead of color selection. The trend of probing a
deeper sky is likely a manifestation of the excellent data provided by
the  SDSS  collaboration  for   the  local  universe.   The  magnitude
selection, on the other hand,  improves the completeness of the survey
by reducing the  cases in which galaxies at  similar distances from us
are  being  dropped  out   of  the  searches  at  optical  wavelengths
\citep[see][and  references   therein]{2005Natur.437..519L}  by  color
selections.

\section{Highlights of Scientific Results}

Some of the science discoveries and results from spectroscopic surveys
are  discussed  in  the  following.   They are  mostly  obtained  upon
analyzing   spectra,  using  line   diagnostics,  template\footnote{By
templates  we mean  model-based or  empirical spectra  for  modeling a
target spectrum.}  fitting or  data mining techniques.  This overview,
if  not  biased, inevitably  omits  tremendous  number of  interesting
results (e.g., currently there are $\approx{1000}$ publications due to
the  SDSS  spectroscopic  data\footnote{By searching  Smithsonian/NASA
Astrophysics Data  System (ADS) using  a title ``SDSS''  or ``Sloan'',
and a keyword ``spectra''  or ``spectroscopy''.}) and even sky surveys
(e.g.,  studies on  large-scale structures  by the  redshifts obtained
from the 2dF redshift  survey).  When appropriate, results obtained by
using photometric or non-survey (i.e., defined as being not one of the
surveys shown in Table~\ref{tab:surveys}) data are also listed.

\subsection{Galaxies}

Presently a main theme in studies of galaxy formation is rooted partly
from  the discovery  of  elliptical galaxies  being  more common  than
spirals   in    galaxy   clusters   \citep{1980ApJ...236..351D},   and
subsequently   the  K$+$A   galaxies   (or  post-starburst   galaxies)
\citep{1983ApJ...270....7D}  in  clusters,  showing that  an  observed
galaxy  spectrum  and  its   stellar  content  depend  highly  on  the
environment.      Up    to    $z     \sim    0.5$,     recent    works
\citep[e.g.,][]{2002MNRAS.333L..31M,2002MNRAS.334..673L,2003ApJ...584..210G,2004MNRAS.353..713K}
showed  that a  high  density environment  (from  clusters to  groups)
suppresses  star  formation  in  its member  galaxies,  with  proposed
reasons include, but are not  confined to, e.g, tidal stripping of gas
away from a  galaxy by its neighbors, and the depletion  of gas in the
galaxies before  the formation of the high  density environment.  This
``quenching''   of   star   formation   is   extended   to   $z   \sim
1$~\citep{2006MNRAS.370..198C}.

Bimodality  in  color-color  and  color-magnitude  diagrams  of  local
galaxies                            is                            seen
\citep[e.g.,][]{2001AJ....122.1861S,2004ApJ...601L..29H},    as   such
early-type  (red)  galaxies  populate   a  narrow  sequence  which  is
separated from  the later-types.  The bimodality and  red sequence are
seen up to $z \sim 1.0$ \citep{2004ApJ...608..752B}.  The formation of
the galaxies in the red sequence may be due to the gas-rich mergers by
blue  galaxies, followed  by  the  dry mergers  by  the resultant  red
galaxies \citep{2005astro.ph..6044F}.

The            MPA/Garching            valued-added            catalog
\citep{2003MNRAS.341...33K,2004MNRAS.351.1151B,2004ApJ...613..898T}
provides high-end physical parameters  of the SDSS DR4 galaxies. Among
many applications, scaling-relationships  such as the mass-(gas-phase)
metallicity relationship are established,  as well as the bimodalities
in the 4000~\AA  \ break and stellar mass of  local galaxies.  The low
yield in  low-mass galaxies (as high as  10$^{10}$~\solarmass) is used
to  argue  against closed-box  model  of  chemical  evolution, and  is
interpreted as a result of metal-enriched galactic wind.

Classification of  star forming  galaxies and Type-II  active galactic
nuclei (AGNs)  using ratios  of emission line-strengths  \citep[or the
BPT  diagram,][]{1981PASP...93....5B}  has been  modified  to a  large
sample of local  galaxies~\citep{2006MNRAS.372..961K}, in which LINERs
are also shown to populate distinct regions from the Type-II AGNs; and
the extension to the  low-metallicity (below solar) narrow-line region
(NLR)  \citep{2006MNRAS.371.1559G}.  On the  other hand,  an automatic
classification     scheme      of     galaxies     is      that     of
Connolly et~al. (1995) by the Karhunen-Lo\`eve (KL) transform (or
the    Principal    Component    Analysis    (PCA),    described    in
\S\ref{sect:class}), which has shown to be applicable to classifying a
large  sample of local  galaxies \citep{2004AJ....128..585Y},  as such
the most important sample variations in the SDSS galaxy spectra are to
manifest  steepness of  the  continuum slopes  and the  post-starburst
activities in galaxies.

Spectral signatures  of Type-Ia supernovae  can be extracted  from PCA
residuals                of               galaxy               spectra
\citep{2003ApJ...599L..33M,2004AAS...205.7104K}.  This  is a promising
way  to quickly  search for  supernovae in  a large  sample  of galaxy
spectra.

A  large population  of  high-redshift ($z  =  1.4 -  5$) galaxies  is
discovered    \citep{2005Natur.437..519L}.    Due   to    the   purely
flux-selected method  ($I$-band), the  sample is several  times larger
than       previous      samples      of       Lyman-break      galaxy
\citep{2004ApJ...604..534S},  in which color  selection is  used.  The
lack of  a large  Ly$\alpha$ flux in  the composite spectrum  of these
galaxies suggests that they may be dusty.

\subsection{Active Galactic Nuclei}

The SDSS quasar (QSO) catalogs \citep[the updated one being the fourth
  version,][]{2007AJ....134..102S}  are unprecedented in  their sample
  sizes (77,429  objects for the  updated catalog).  The  selection of
  majority  of   the  QSOs  from   the  multi-dimensional  color-space
  \citep{2002AJ....123.2945R} for follow-up  spectroscopy results in a
  high completeness and efficiency.

The  discoveries  of  high-redshift  QSOs  at far  as  $z  \approx  6$
\citep{2001AJ....122.2833F} have made impacts in both galaxy formation
and  cosmology.   Particularly,  the  detection of  the  Gunn-Peterson
trough  (the zero monochromatic  flux shortward  of Ly$\alpha$  due to
absorptions by intergalactic neutral hydrogen) in a $z=6.28$ QSO and a
non-detection  in  a  $z  \approx 6$  one  \citep{2001AJ....122.2850B}
suggests that hydrogen in the universe has undergone a transition from
a neutral to an ionized state at redshift of $\approx6$.

Type-II   (i.e.,    narrow   emission-line)   QSOs    are   discovered
\citep{2003AJ....126.2125Z},  which are  luminous AGNs  as  such their
central regions are shielded by large amounts of gas and dust. Before,
only broad emission-line luminous AGNs were known.

QSO spectra display a large diversity, which is shown to be related to
the effects of redshift (more prominent iron emissions in low-$z$ QSOs
to   the  high-$z$   ones)   and  luminosity   (the  Baldwin   effect)
\citep{2004AJ....128.2603Y}.  Further, some broad absorption line QSOs
(BALQSOs)  show  unusual  spectra  \citep{2002ApJS..141..267H},  which
require  various explanations  such as  a particular  geometry  in the
outflow  materials to absorb  the continuum  radiation.  On  the other
hand, Boroson (2002)  proposed that the black hole  mass and Eddington
ratio  drive the properties  seen in  both radio-loud  and radio-quiet
QSOs.

Spectral        decomposition         of        AGN-host        galaxy
\citep{2004AJ....128.2603Y,2006AJ....131...84V}  by  the  KL-transform
provides a new way to  separate the host galaxy contributions from the
otherwise host-contaminated Type-I AGN spectrum, thus serves as a tool
to    study   the    co-evolution   between    AGN   and    its   host
galaxy. Post-starburst  activities are found  to be more  common among
AGN host galaxies than inactive galaxies \citep{2006AJ....131...84V}.

High-ionization broad-emission  lines, such as  C~IV, are found  to be
blueshifted   ($\lsim   2,000$~km~s$^{-1}$)   with  respect   to   the
low-ionization  broad-emission  lines   (e.g.,  Mg~II)  in  some  QSOs
\citep{2002AJ....124....1R}.  The behavior  led the authors to suggest
that orientation,  whether external or  internal, may be the  cause of
the effect.

Black hole  mass (estimated from broad-emission  line kinematics, with
the  assumption of  the virial  radius from  results  on reverberation
mapping)   of    the   SDSS   QSOs    shows   an   upper    limit   of
$3\times{10}^{9}$~\solarmass~\citep{2004MNRAS.352.1390M},    consistent
with the mass of the largest black holes found in the local universe.

The extinction  curves of QSOs were  found to be similar  to the Small
Magellanic Cloud  (SMC) extinction curve  with a rising  UV extinction
below   2200~\AA~\citep{2006MNRAS.367..945Y}.   This  adds   one  more
supporting case for the unique  extinction properties of the Milky Way
galaxy, in which the 2175~\AA  \ bump (a feature often associated with
graphite grains) is present.

\subsection{Stars}

The recent  discoveries of 10 new  satellites of the  Milky Way galaxy
\citep[see][and   references  therein]{2007ApJ...654..897B}  certainly
contribute  to  one of  the  biggest impacts  by  the  SDSS. They  are
expected to attract  many studies in near-field cosmology,  as well as
in implications on hierarchical galaxy formation as a whole.

By modeling  the kinematics of  high velocity stars, the  local escape
velocity of the  Milky Way galaxy is found  to be significantly larger
then   the   local   circular  velocity   \citep{2007IAUS..235..137S},
suggesting  the  existence  of  mass  exterior to  the  Solar  circle,
possibly the dark-matter halo.

\section{Spectral Analyses}

\subsection{Model-based vs. Empirical Approach}

Spectral analyses  in the model-based  approach, e.g., the  fitting of
galaxy    spectra   with    stellar   population    synthesis   models
\citep[e.g.,][]{1997A&A...326..950F,1999ApJS..123....3L,2003MNRAS.344.1000B}
and of quasar  spectra with theoretical \citep{2003ApJS..145...15S} or
empirical            templates            of            iron-emissions
\citep{2001ApJS..134....1V,2004A&A...417..515V},  make the extractions
of physical  parameters straightforward, as the  physics is prescribed
by   the  model.   However,   they  are   hindered  by   the  possible
imperfections in models, such as the incompleteness of ingredients and
the incorrect  assumptions of physical conditions.   Besides, they may
not be the best approach in making discovery about the data.

On the contrary, without  a priori descriptions the empirical approach
is more favorable  to making discovery in the  data.  For example, the
constructions of  composite spectra or  KL eigenspectra from  the data
belong  to  this  category.   With  the lack  of  model  descriptions,
however,  interpretations  are   usually  needed  for  the  underlying
physics. One  may also need to  resort to models  later for estimating
physical parameters.

In  these  regards,  none  of  the  approaches  processes  substantial
advantage over the  other, thus both could be  used in a complementary
fashion when studying a specific problem.

\subsection{Minimization Techniques}\label{sect:fitting}

Popular  techniques  used in  astronomy  for  minimizing spectra  with
templates are  shown in Table~\ref{tab:fitting}.   For future analysis
on  complex models  (e.g., as  much as  $\approx10$ parameters  in the
P\'EGASE  stellar population  model; or  a large  number  of fitting
components  --  emission and  absorption  lines  of different  widths,
shapes and possible shifts  from laboratory wavelengths), the Bayesian
approach  is  promising.   As  it  has a  more  logical  and  coherent
framework in characterizing the likelihood surface defined by the data
and the model, which may be degenerate.

To  extract  quickly  physical   parameters  from  galaxy  spectra,  a
promising method  is the  Multiple Optimised Parameter  Estimation and
Data  compression   \citep[MOPED,][]{2000MNRAS.317..965H},  a  maximum
likelihood method  which reduces  the number of  wavelength bins  in a
spectra   without  increasing   the  uncertainty   in   the  estimated
parameters. It  has been  applied to the  fast estimation of  the star
formation history,  metallicity and dust content  of galaxies \citep[a
few seconds for 17 parameters,][]{2001MNRAS.327..849R}. At present, it
is  unknown  for  the  scopes  of  probability  distribution  (or  the
complexity of the model) where this method is applicable to. 

\subsection{Statistical Classifications}\label{sect:class}

Classification  of  spectra  were  traditionally carried  out  through
visual inspection,  often accomplished by  a small number  of experts.
New  methods are  inevitable for  studying large  sets of  data, which
should  be  able  to  consider relevant  information  objectively  and
automatically, and to reduce the error rate.

One  of   the  first   works  is  by   Connolly  and   his  co-workers
\citep{1995AJ....110.1071C}  in  which  a  linear  combination  of  an
orthogonal  set   eigen-functions  (called  ``eigenspectra''   by  the
authors) was  adopted to represent  each galaxy spectrum.   The method
used to  construct the eigenspectra, namely  the Karhunen-Lo\`eve (KL)
transform, is a powerful  and popular technique used in classification
and dimensional reduction of massive data sets.  The eigenspectra were
shown by  the authors to  be applicable to classifying  galaxy spectra
into  a  sequence  ranging  from   the  old  to  the  younger  stellar
populations  (later known as  the ``eClass'').   The KL  transform has
been   applied   in  the   spectroscopic   classifications  of   stars
\citep{1998MNRAS.298..361B,1998MNRAS.295..312S},               galaxies
\citep{1995AJ....110.1071C,1997ASPC..114..149S,1998ApJ...505...25B,
1998A&A...332..459G,1999MNRAS.303..284R,1999MNRAS.308..459F,
2004AJ....128..585Y}                      and                     QSOs
\citep{1992ApJS...80..109B,1992ApJ...398..476F,
2003ApJ...586...52S,2004AJ....128.2603Y}.    This   method  has   been
extended  to spectra  which are  ``gappy'', e.g.,  with bad  pixels or
skylines \citep{1999AJ....117.2052C}.

The  other notable  method used  in classification  is  the artificial
neural network.  Also called  ``supervised learning'', the idea behind
is the users manually classify a subset of the data, and from this the
neural network is  ``trained'' to classify the remaining  of the data.
This method  has been applied in the  spectroscopic classifications of
stars
\citep{1994MNRAS.269...97V,1997ApJ...487..847W,1997PASP..109..932B,
1998MNRAS.295..312S,2001ApJ...562..528S},                      galaxies
\citep{1996MNRAS.283..651F} and AGNs \citep{1996PASA...13..207R}.

There  are many works  applying other  machine learning  techniques to
classification  problems in  astronomy.  The  decision  tree induction
(returns  a boolean  function  ``yes'' or  ``no''  by finding  logical
patterns  within  the  input   data)  has  been  used  in  star-galaxy
separation   \citep{1995AJ....109.2401W,2004AJ....128.3092O}   and  in
morphological classification  of galaxies \citep{1996MNRAS.281..153O}.
The information  bottleneck method  (finds the best  trade-off between
accuracy and  compression when clustering  a random variable  x, given
the joint probability distribution  between x and an observed variable
y)    has   been   applied    to   galaxy    spectral   classification
\citep{2001MNRAS.323..270S}.

A recent addition is a PCA-based approach which probes the correlation
between nebular  emissions and stellar  continua of the  SDSS galaxies
\citep{eigenlines}, in  the subspace spanned  by the eigencoefficients
(i.e.   the  expansion  coefficients  of  a spectrum  by  the  set  of
eigenspectra) of both  the emission lines and the  continua.  This may
be  a  promising  approach   for  estimating  statistically  the  line
strengths given a stellar continuum.

\section{Present Needs \& Challenges}

\subsection{Modeling Galaxy}

The spectrophotometry is  unlikely to be perfect in  both the observed
and the template spectra from stellar population models.  As a result,
a    true   spectrum   $f_{\lambda}$    is   changed    generally   to
$f^{\prime}_{\lambda}$, as such

\begin{equation}
f_{\lambda} \rightarrow f^{\prime}_{\lambda} = g_{\lambda} f_{\lambda} \label{eqn:spectrophoto}
\end{equation}

where    $g_{\lambda}$   is    a    wavelength-dependent   array    of
values.    Line-strength   indices    that   are    represented   by
equivalent-width (EW) are not affected, as 

\begin{equation}
{\rm    EW}^{\prime}   \equiv   \int    \frac{f^{\prime}_{\lambda}   -
c^{\prime}_{\lambda}}{c^{\prime}_{\lambda}}   d\lambda   \stackrel{\rm
by~Eqn.~\ref{eqn:spectrophoto}}{=}     \int     \frac{f_{\lambda}    -
c_{\lambda}}{c_{\lambda}} d\lambda = {\rm EW} \ ,
\end{equation}

\noindent 
assuming the  continnum flux densities  $c^{\prime}_{\lambda}$ is also
affected according to  Eqn.~\ref{eqn:spectrophoto}, a valid assumption
for  atmospheric  differential  refraction  or  instrumental  effects.
Other  line  diagnostics  can  be  checked accordingly  in  a  similar
fashion.  However,  derived quantities are affected if  they depend on
the  absolute flux  scale of  the spectrum,  e.g., stellar  mass  of a
galaxy  from  fitting  stellar  population  models  to  the  spectrum,
synthetic magnitudes of the spectrum.

\subsection{Parameterization of Star Formation History}

The optimal parameterization of the star formation history of galaxies
is unknown.  How many parameters at maximal can we fit for an observed
spectrum by  templates (in this  context, a mixture of  simple stellar
populations of  arbitrarily different  ages and metallicities,  and an
intrinsic dust reddening model) without  starting to fit for the noise
in the  spectral flux densities?  A possible  approach to characterize
this is to perform tests on parameter recovery: starting from a simple
model, and subsequently adding more complexity (e.g., by relaxing more
parameters to change).

\subsection{Modeling Quasar}

Quasar spectra present a challenge  to modelers for a general approach
because  of  their  displayed  diversity.   Line-widths,  as  well  as
line-shapes,  can  be different  on  both  the element-to-element  and
object-to-object  bases.   There  are  high-ionization  broad-emission
lines    that     may    be    blueshifted    for     as    much    as
$\approx2,000$~km~s$^{-1}$ \citep{2002AJ....124....1R}.  Further, pure
emission and absorption lines are  difficult to be isolated because of
the ``contamination''  by iron  emissions (as much  as $\approx20,000$
transitions in Fe~II from  UV to IR \citep{2003ApJS..145...15S}, which
at times  form a pseudo-continuum), prominent effects  in the vicinity
of  Mg~II and  H$\beta$ emissions  for some  QSOs. For  BALQSOs, their
features are found  to span a number of  higher-order eigenspectra and
are     not     confined    to     only     one    particular     mode
\citep{2004AJ....128.2603Y}.  All  of these factors  contribute to the
fact that a  compact description of QSO spectra  (e.g., with a handful
of eigenspectra) is unlikely.

\subsection{Cross-Calibration}

While cross-wavelength, -catalog and -epoch analyses often provide new
insights to  science problems,  there is no  scheme yet  available for
calibrating  generally the  datasets from  different  categories.  For
example, it is not clear  how to separate systematics from statistical
errors,  where both  of them  can be  completely different  across the
catalogs.   This will  undoubtedly  pose a  challenge  in designing  a
general  data model  for recording  a priori  the  relevant parameters
which  are   useful  for  hunting   down  the  systematics,   such  as
observational conditions or instrumental characteristics.

\subsection{Parameter Estimations}

Thus  far,  the  uncertainties  of estimated  physical  parameters  of
spectra quoted in literatures are usually obtained by the bootstrap or
similar  method  (with  the   1$\sigma$  \  uncertainties  being  most
popular). However, it is not  proved that such an uncertainty is equal
to that  obtained from error propagations of  parameter uncertainty of
each object in the  sample, particularly under different situations of
realistic S/N where the  underlying likelihood surface (defined by the
data and model) is expected to be modulated accordingly.

\section{Spectroscopic Science \& Virtual Observatory}

The greatest benefit astronomers gain from using the tools provided by
the VO \citep{2001Sci...293.2037S} is probably the unification of data
access.  For spectroscopic  data, the standard is a  XML Schema called
the  Spectrum Data  Model,  established by  the International  Virtual
Observatory  Alliance  (IVOA).   Such  standards  are  invaluable  for
speeding up  error-checking and  peer communications. An  unified data
access is  hereby made  possible by the  VO Registry,  eliminating the
need for storing locally a copy of the data by each user.

In  addition, the developed  tools in  archiving spectra  and spectral
analyses   are   readily  available   to   the  astronomy   community,
particularly   in  archiving   survey  catalogs   \citep[NVO  Spectrum
Services,][]{2004ASPC..314..185D},      multi-wavelength      analyses
\citep[VOSpec,][]{2005ASPC..347..198O},       and       visualizations
\citep[Specview][]{2000ASPC..216...79B}.  Further, all of them provide
some common  spectral analyses, and  fitting services to  spectra with
templates.   On  the  other  hand, the  searching  and  cross-matching
services  for multiple  tables  or catalogs,  e.g.,  the SDSS  CasJobs
\citep{2005cs........2072O} and  the WESIX \citep{2005ASPC..347..355K}
would  be  very  useful  for  spectral  analyses  in  multi-epoch  and
-wavelengths.

A  checklist of  the needs  for  astronomers in  their daily  research
likely includes

{\begin{small}
\begin{itemize}
\item unique sets of data
\item easy uploading of user data 
\item a complete inventory for searching available datasets
\item accurate documentations of  schema's and web services
\item an unique set of standard tools for common spectral analyses such as foreground
  dust dereddening, constructions of composite spectra, and simple fitting tasks
\item being notified of version updates
\end{itemize}
\end{small}
}

\section{Acknowledgments}

I thank the organizers of the Astronomical Spectroscopy \& The Virtual
Observatory WorkShop  2007 for inviting  me to attend the  meeting and
giving  this talk.   I  thank Tam\'as~Budav\'ari,  S\'ebastien~Heinis,
Rosemary~F.~G.~Wyse  and   Alex~Szalay  for  useful   discussions.   I
acknowledge support through grants  from the W.M.  Keck Foundation and
the  Gordon and  Betty Moore  Foundation,  to establish  a program  of
data-intensive science at the Johns Hopkins University.

\bibliographystyle{vospwks2007}

\clearpage

\begin{table}
\caption{Some of the current spectroscopic surveys.} 
\begin{tabular}{llrrll}
\hline
Survey         & Year & $N${${^a}$}       & $R${${^b}$} & $z$ & $m_{\rm Lim}${${^c}$} \\
\hline
SDSS-I{${^d}$} & 2000 & 10$^6$            & 2000 & 0 $-$ 0.3   & $r < 17.77$${^e}$ \\
               &      & 10$^5$            & 2000 & 0 $-$ 6     & $r < 19.1$${^f}$  \\
SDSS-II{${^g}$}& 2005 & 2$\times10^{5}${${^h}$}   & 2000 & 0 $-$ 6     & same as above  \\
DEEP2{${^i}$}          & 2002 & 6$\times10^{4}$   & 5000 & 0.7 $-$ 1.5 & ${I_{\rm AB}} < 24.5$   \\
VVDS{${^j}$}   & 2002 & $10^{5}$          & 250  & 0.2 $-$ 4.0 & ${I_{\rm AB}} < 22.5$ \\
               &      & 5$\times10^{4}$   & 250  & 0.2 $-$ 4.0 & ${I_{\rm AB}} < 24$   \\
               &      & $10^{3}$          & 250  & 0.2 $-$ 4.0 & ${I_{\rm AB}} < 26$ \\
zCOSMOS{${^k}$}& 2005 & 2$\times10^{4}$   & 600  & 0.1 $-$ 1.2 & ${I_{\rm AB}} < 22.5$ \\
               &      &        $10^{4}$   & 200  & 1.4 $-$ 3.0 & color{${^l}$} \\
RAVE{${^m}$}   & 2006 &        $10^{6}$   & 7500 & Galactic    & $9 < {I} < 12$  \\
\hline
\end{tabular}
{\begin{small}
{$a.$ Number of spectra.}  \\
{$b.$ Spectral resolution, $R = \lambda/\Delta\lambda$.} \\
{$c.$ Magnitude limit(s), for which an object with brightness higher than (or within) are included in the survey.} \\
{$d.$ The  Sloan  Digital  Sky  Survey  \citep{2000AJ....120.1579Y}  is
  currently  the  largest sky  survey  with  both fiber-fed  
  spectroscopy and $ugriz$ photometry.} \\
{$e.$ The galaxy survey, with a completeness  $>99\%$~\citep{2002AJ....124.1810S}. } \\
{$f.$ The quasar (QSO) survey, with a completeness  $\approx89\%$~\citep{2005AJ....129.2047V}.} \\
{$g.$ The SDSS-II comprises three surveys: the Legacy, the Sloan
  Extension for Galactic Understanding and Exploration (SEGUE) and
  the Supernovae surveys.} \\
{$h.$ The projected number of spectra respectively for both the Legacy
and the SEGUE surveys ({c.f., Kron~R., in http://www.sdss.org/surveyops/SDSS-II/nsf$\_$First$\_$Year$\_$Review/nsf06-overview.pdf})}. \\
{$i.$ The DEEP Extragalactic Evolutionary Probe 2 Redshift Survey \citep{2003SPIE.4834..161D}.}\\ 
{$j.$ The VIMOS VLT Deep Survey \citep{2004A&A...428.1043L}.} \\
{$k.$  The   redshift  survey \citep{2006astro.ph.12291L}  in  the
  COSMOS field \citep{2006astro.ph.12306S} comprises zCOSMOS-bright
  and -deep, with complementary imaging data from X-ray to radio.}\\
{$l.$ This includes  the BzK selection for searching star-forming
  and passively-evolving galaxies in $1.5 < z < 2.5$, and possibly other color selections;
  with an additional magnitude cut of ${B_{\rm AB}} < 25$ for the desired
  continuum S/N.} \\
{$m.$ The RAdial Velocity Experiment \citep{2003ASPC..298..381S}. Spectra are taken in
  $\lambda\lambda 8410-9795$~\AA, which includes the Ca-triplet.}\\
\end{small}}
\label{tab:surveys}
\end{table}

\clearpage

\begin{table}
\caption{The commonly used minimization techniques of spectra and
  their main characteristics.} 
\begin{tabular}{lll}
\hline
Method & Advantage & Dis-advantage \\
\hline
             least-square fit & -- simple implementation & -- works well for
	     Gaussian \\
& & probability distribution only \\
non-negative least-square fit{$^{a}$} & -- each monochromatic flux
from template is
summed,  & -- still least-square fit\\
  & positively, hence more physical than least-square fit  & \\
   Bayesian approach & -- logical and coherent framework for uncertainty    & -- may need self implementation \\
\hline
Scheme & Advantage & Dis-advantage \\
\hline
line indices{$^{b}$} & -- focus on, e.g., age- and metallicity-sensitive lines
& -- depend on prior knowledge of lines \\
whole spectrum & --  contributions from all wavelength bins  & -- irrelevant lines may
interfere \\
& & --  still model-dependent \\
\hline
\end{tabular}
{$a.$ See, e.g., \cite{NNLS}. There are also publicly available
  softwares in C, Fortran, idl and Matlab.} \\
{$b.$ See, e.g., the Lick system \citep{1998ApJS..116....1T}.}
\label{tab:fitting}
\end{table}

\end{document}